\documentclass{IEEEtran}
\usepackage{cite}
\usepackage{amsmath,amssymb,amsfonts}
\usepackage{graphicx}
\usepackage{textcomp,nicefrac}
\def\BibTeX{{\rm B\kern-.05em{\sc i\kern-.025em b}\kern-.08emT\kern-.1667em\lower.7ex\hbox{E}\kern-.125emX}}
\usepackage{makecell}
\usepackage{subfigure}

\begin{document}
\title{Test to the Performance of a LGAD Based Zero Degree Detector on CSNS Back-n Beamline}
\author{Yuhang Guo, 
        Mengzhao Li, 
        Mengchen Niu,
        Zhijun Liang,
        and Ruirui Fan
%\thanks{This paragraph of the first footnote will contain the date on which you submitted your paper for review. It will also contain support information, including sponsor and financial support acknowledgment. For example, ``This work was supported in part by the U.S. Department of Commerce under Grant BS123456.'' The next few paragraphs should contain the authors' current affiliations, including current address and e-mail. For example, F. A. Author is with the National Institute of Standards and Technology, Boulder, CO 80305 USA (e-mail: author@ boulder.nist.gov).}
\thanks{This work was supported in part by the China Postdoctoral Science Foundation under Grant 2023M732034.}
\thanks{Y. Guo is with Spallation Neutron Source Science Center, Dongguan 523803, China, also with Institute of High Energy Physics, CAS, Beijing 10049, China.}
\thanks{M. Li is with Institute of High Energy Physics, CAS, Beijing 10049, China, also with China Center of Advanced Science and Technology, Beijing 100190, China.}
\thanks{Z. Liang is with Institute of High Energy Physics, CAS, Beijing 10049, China.}
\thanks{M. Niu is with Spallation Neutron Source Science Center, Dongguan 523803, China.}
\thanks{R. Fan is with Spallation Neutron Source Science Center, Dongguan 523803, China, and with Institute of High Energy Physics, CAS, Beijing 10049, China, also with State Key Laboratory of Particle Detection and Electronics, Beijing 100049, China. (e-mail: fanrr@ihep.ac.cn)}
}

\maketitle

\begin{abstract}
A comprehensive and reliable nuclear database of neutron-induced process is required by the development of Accelerator-Driven System technology.
Although many excellent efforts have been made, the measurements to the fission fragment in zero degree of the beam direction is not perfect enough. 
Most detector can not work constantly under the strike of intense flux neutron beam, thus their application as Zero Degree Detector (ZDD) is greatly limited.
%No detector is available for this application as it has to work steadily under the strike of neutron beam.
Recently, the Low Gain Avalanche Diode (LGAD) technology has been developed to meet the complicated and several requirement of ATLAS on Large Hadron Collider (LHC) and could survive after $\mathbf{2.5\times 10^{15}~n_{eq}/cm^2}$ irradiation.
With its 50 \textmu m active thickness, 30 ps temporal resolution, low cost and good radiation hardness property, the LGAD detector is expected to be an excellent candidate for the use of a Zero Degree Detector in neutron-induced process data measurements. 
This paper represents a performance test to LGAD detector when it is used as a ZDD on the Back-streaming Neutron (Back-n) beamline at CSNS. 
Measuring to the cross section of $\mathbf{^6Li(n,T) \alpha}$, it shows that the LGAD performs well as a ZDD based on a good agreement with the data from ENDF database. 
\end{abstract}

\begin{IEEEkeywords}
TBD
%Enter key words or phrases in alphabetical order, separated by commas. For a list of suggested keywords, send a blank e-mail to keywords@ieee.org or visit \underline{http://www.ieee.org/organizations/pubs/ani\_prod/keywrd98.txt}
\end{IEEEkeywords}

\section{Introduction}
\label{sec:intro}

Accelerator-Driven Systems (ADS) is one of the promising technology for future nuclear power plants. 
It will not only enhance the safety measures for controlling conventional nuclear reactors but also improve the efficiency of nuclear fuel utilization. 
However, the development of ADS technology necessitates a comprehensive and reliable nuclear database of neutron-induced processes, which is yet to be fully achieved and requires significant efforts. \cite{ADS-tb}

The cross section and the angular distributions of the fragments are important issues of neutron induced process data measurement.
Many beneficial works have been done to measure the angular distributions, with gaseous or silicon detector series setup. \cite{NDM-gas-1,NDM-gas-2,NDM-gas-3,NDM-gas-4,NDM-Si-1,NDM-Si-2}
However, the measurement of fragments distributions in zero degree direction of the neutron beam direction is a challenging work.
This is primarily due to the direct irradiation of the zero-degree detectors (ZDD) by the neutron beam. 
Gaseous detectors, in this scenario, are easily interrupted by sparks, while silicon detectors are bothered by the irradiation damage. 

Recent studies have focused on exploring new types of detectors that could potentially be used for ZDD. 
SiC (silicon carbide) detectors \cite{SiC} and diamond detectors \cite{Diamond} are considered to be promising candidates due to their excellent anti-irradiation properties because of their wide forbidden band gaps. 
However, the application of SiC detectors is limited by their small surface area and large thickness, while diamond detectors are too expensive for widespread use. 
Further research is needed to overcome these limitations and find cost-effective solutions for implementing ZDD in neutron-induced process measurements.

In recent years, many advanced detector technologies are being developed to meet the high-performance requirements of the Large Hadron Collider (LHC). \cite{LGAD-1}
The Low Gain Avalanche Diode (LGAD) \cite{LGAD} technology is one of them that is particularly well-suited for the ZDD applications. 
The LGAD detector is N-on-P silicon sensor fabricated on 50~\textmu m epitaxial layer of a p-type silicon bulk.
A moderate internal gain, typically ranging from 10 to 30 is achieved by establishing high electric field within the multiplication layer between the p-n junction, and consequently enhances the temporal resolution to 30~ps approximately. 
LGAD detectors offer excellent timing resolution and radiation hardness, making them highly suitable for precise measurements in high-energy physics experiments. \cite{LGAD-irr}
Table \ref{tab:lgad} lists some important performance parameters of LGAD.

\begin{table}[h]
\centering
\caption{Selected performance parameters of LGAD detector. \cite{LGAD,LGAD-1}} 
\begin{tabular}{c|c}
\Xhline{1.2pt}
Performance              & Parameters           \\
\Xhline{1.2pt}
segmentation dimensions & $1.3\times 1.3~\mathrm{mm^2}$ \\
\hline
active thickness          & 50 \textmu m           \\
\hline
temporal resolution      & ~30 ps                \\
\hline
gain                     & 10$\sim$30           \\
\Xhline{1.2pt}
\end{tabular}
\label{tab:lgad}
\end{table}

The active thickness of LGAD is 50~\textmu m. 
Neutrons and $\gamma$ have a minimal tendency to deposit energy within the sensitive volume of a detector, whereas ionized particles have the ability to fully deposit their energy.
Additionally, it can be produced as a detector array with a pixel size of  about $1\times 1~\mathrm{mm^2}$, which enables the simultaneous detection of both light and heavy particles.
Furthermore, the LGAD detector offers a temporal resolution of approximately 30~ps, which greatly benefits the measurement of neutron energy using the time-of-flight method. 
With these advantages, there is a promising potential for the LGAD detector to be utilized as the ZDD in the measurement of the angular distribution of neutron-induced process fragments.

% It will be better if there is a table of the comparision of the detectors.

In order to check this idea, experimental tests were conducted on the Back-n beamline at the China Spallation Neutron Source (CSNS).
This paper is organized as the followings.
The experimental setup is introduced in Sec.\ref{sec:setup}, including the general properties of CSNS and Back-n, a preliminary ZDD setup and the corresponding DAQ electronics.
The analysis to the data collected and the results are introduced in the subsequent section.
Finally, the conclusion will be presented in the last section, as well as the insights into the potential applications of the LGAD detector as a ZDD for measuring the angular distribution of neutron-induced charged particles.

\section{Experimental Setup}
\label{sec:setup}

\subsection{Back-n Beamline}

\begin{figure}[t]
\centering
\includegraphics[width=.5\textwidth]{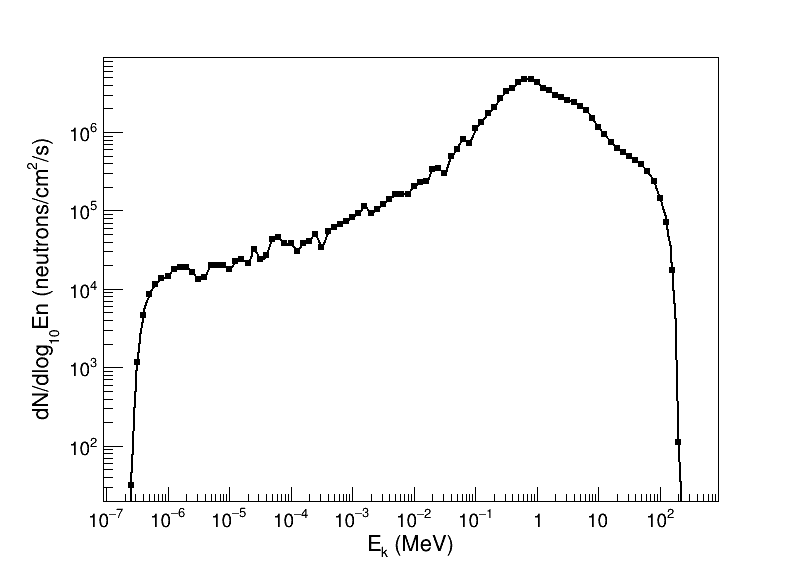}
\caption{Kinetic energy spectrum of the neutron provided by Back-n, spanning from thermal neutron energy to several hundreds of MeV. \cite{Ek}\label{fig:Ek}}
\end{figure}

The Back-n is a white neutron beamline that has been constructed within the China Spallation Neutron Source (CSNS). 
Its primary purpose is to facilitate multidisciplinary research in material science through the utilization of neutron scattering methods.
The accelerator operates in a double bunch mode, where two clusters of protons are generated with a time interval of 410~ns. 
These protons are then accelerated to a energy of 1.6 GeV within the Rapid Cycling Synchrotron (RCS). 
Subsequently, they collide with a tungsten target at a repetition frequency of 25~Hz.
The designed beam power of CSNS in phase-I is 100~kW, which would be upgraded to 500~kW in phase-II.
As a result of the spallation reaction, neutrons are produced and scattered in all directions, covering a $4\pi$ solid angle. 
The Back-n takes advantage of the neutrons that back-stream in the direction of the incoming protons.
As illustrated in Figure \ref{fig:Ek}, the Back-n beamline offers a wide range of kinetic energy for neutrons, spanning from thermal neutron energy to several hundreds of MeV. 
Additionally, the flux density of the Back-n is $7.03\times 10^{6}~\mathrm{neutrons/cm^{2}/s}$ at 100~kW.
This characteristic makes it highly suitable for measuring neutron-induced process data. 
For a more comprehensive understanding of Back-n, please refer to \cite{WNS}.

\subsection{The Detector Setup}

%A ZDD setup is built 

\begin{figure}[h]
    \centering
    \subfigure[Photo of the detector on PCB]{    
		\label{fig_s}     
	    \includegraphics[width=0.4\textwidth]{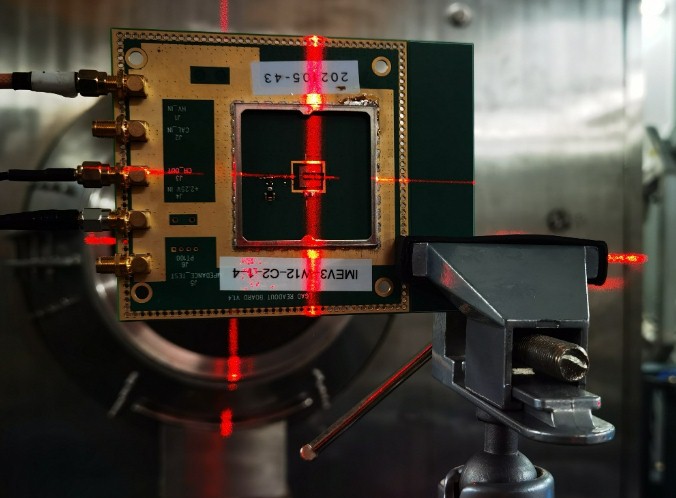}}
    \qquad
    \subfigure[The detector cover and the lithium target]{    
		\label{fig_t}     
		\includegraphics[width=0.4\textwidth]{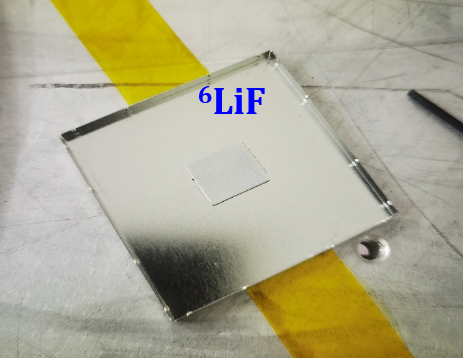}}
	\caption{Photos of the detector setup. The LGAD detector on PCB (a) is fixed on the center of back-n beamline with the help of laser indicators. The detector is covered by a aluminum cover (b), on which a lithium target is attached on it.}
	\label{fig:det}
\end{figure}

As illustrated in the top plot of Figure \ref{fig:det}, a LGAD detector in size of $1.3\times 1.3~\mathrm{mm^2}$ is bonded onto a printed circuit board (PCB) along with a pre-amplifier. 
The detector applied is in batch of IHEP-IME, which is designed and fabricated independently by the Institute of High Energy Physics, CAS, China. \cite{LGAD-irr}
The signals generated by the detector are pre-amplified and subsequently read by the readout electronics, which will be discussed in detail later in the paper.
To shield the detector from ambient light, an aluminum cover is placed on top of the PCB. 
Additionally, as illustrated in the bottom plot of Figure \ref{fig:det}, a Lithium target is attached to the center of the cover, directly above the detector. 
The Lithium target consists of a 50~\textmu m layer of $^6LiF$ plated on a 0.5~mm thick aluminum substrate.
The detector is positioned closely to the target, with a separating distance of approximately 1~mm. 
For precise alignment, laser indicators are employed to guide the installation process, ensuring that the LGAD detector is accurately centered on the neutron beam. 
The distance between the detector and the center of the spallation target is measured to be 77.0~meters.

\begin{figure}[ht]
\centering
\includegraphics[width=.48\textwidth]{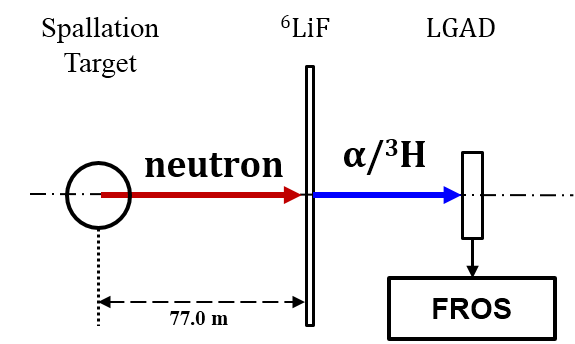}
\caption{Experimental scheme of the detector setup for $^6Li(n,T)\alpha$. (Not to scale)\label{fig:scheme}}
\end{figure}

The experiment is illustrated in Figure \ref{fig:scheme}. 
As the neutrons traverse the lithium target, $^6LiF(n,T)\alpha$ will occurs in possibilities, resulting in the production of $\alpha$ and tritium in $4\pi$ solid angle.
These secondary particles are then detected if they hit on the LGAD detector. 
The signals from the detector are read by the electronics FROS, which will be introduced in later section.

\subsection{The Readout System}

\subsubsection{The Fast Read Out System}

The LGAD detector is known for its high timing resolution capabilities. 
Its signal exhibits a rising time of approximately 700~ps and a full width of 2.5~ns. 
To effectively read such a fast signal, the readout electronics must be characterized by a high sampling rate and bandwidth.
In this experiment, a Fast Read Out System (FROS) is employed to meet these requirements. 
The FROS is a high-speed readout system based on the PXIe protocol, as illustrated in Figure \ref{fig:FROS}. 
The acquisition card used in the FROS setup is the PXIe-X1022, manufactured by Hefei EverACQ Technologies Co., Ltd.
Each acquisition card is responsible for reading signals from a single channel, with a sampling rate of 6.4~Gsps and a bandwidth of 2~GHz. 
These cards are carried by a PXIe crate. 
The acquired data is digitized and then summarized by the controller PC of the crate.
The FROS system is capable of supporting signals readout from one to 17 channels simultaneously. 

\begin{figure}[h]
    \centering
    \subfigure[FROS and power suppliers]{    
		\label{fig_s}     
	    \includegraphics[width=.4\textwidth]{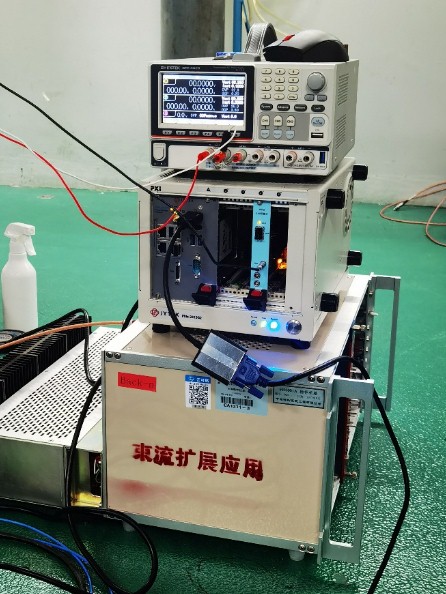}}
    \quad
    \subfigure[Close-up photo of FROS]{    
		\label{fig_t}     
		\includegraphics[width=.4\textwidth]{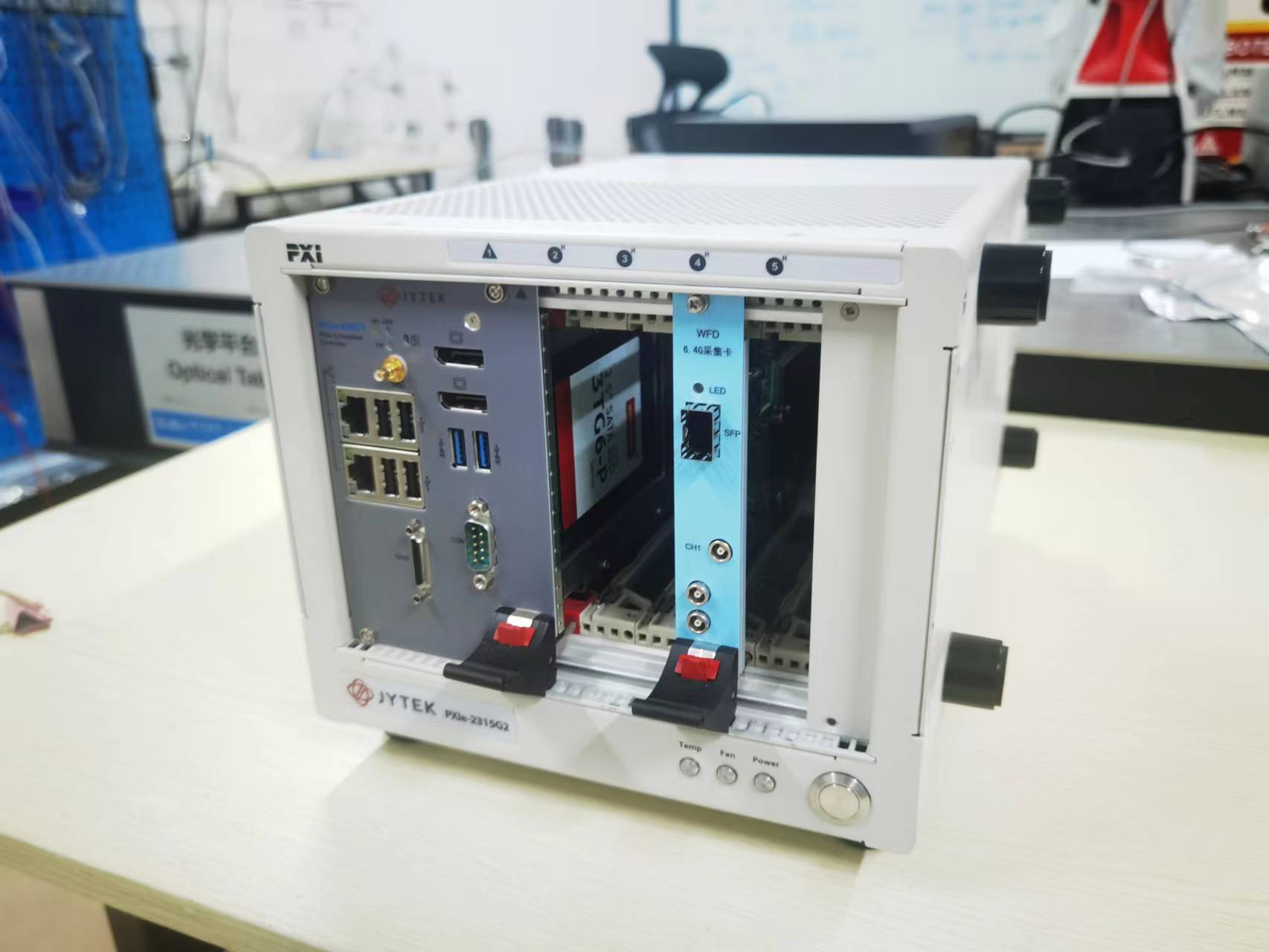}}
	\caption{Photo of FROS and the power supply devices. In (a), the middle device is the FROS, while the top and bottom devices are low and high voltage power suppliers. (b) is the close-up of FROS. The signals and the $T_{0}$ signals are accessed with LEMO connectors. \label{fig:FROS}}
	\label{fig:FROS}
\end{figure}

To ensure synchronization, the clocks of the cards within the FROS are synchronized using a timer trigger card.
A 25~Hz pulsed signal, referred to as the $\mathrm{T_0}$ signal, serves as a time reference. 
This signal is generated by the RCS kicker when the protons are extracted from the RCS. 
The $\mathrm{T_0}$ signal is accessible to the acquisition card, allowing for the recording of time differences between the sampling signals and the $\mathrm{T_0}$ signal.

\subsubsection{Function Validation to FROS}

\begin{figure}[h]
\centering
\includegraphics[width=.48\textwidth]{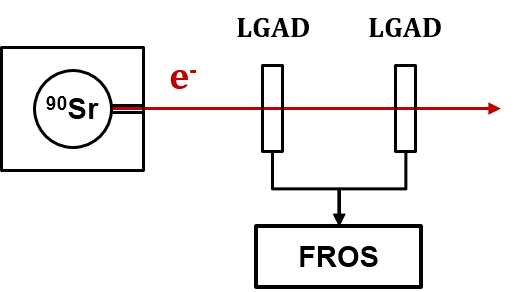}
\caption{Scheme of the function validation to the FROS. Two LGAD detectors are placed in line to measure the collimated electrons from the $^{90}Sr$ source. \label{fig:valid-scheme}}
\end{figure}

The performance of FROS is functionally validated as illustrated in Figure \ref{fig:valid-scheme}. Two LGAD detectors and a $^{90}Sr$ source are deployed in line. 
The electrons released from the source are collimated by the Lead enclosure of the source and directed to the LGAD detectors.
Most of the electrons could go through the first LGAD and reach the second LGAD since they are in minimal ionization particle state. 
The signals from the both LGAD are expected to be recorded with a time difference, because their distances from source are different.
This time difference, which is actually the time of electron flight ($TOF_e$), can be utilized to evaluate the temporal resolution of the LGAD.

\begin{figure}[h]
    \centering
    \subfigure[Sample of a signal read by FROS]{    
		\label{fig_s}     
	    \includegraphics[width=0.49\textwidth]{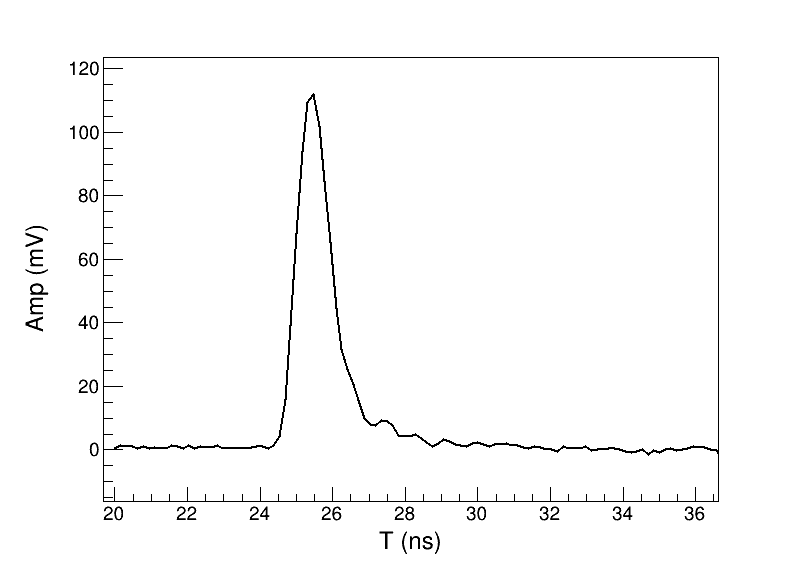}}
    \subfigure[Distribution of $TOF_e$]{    
		\label{fig_t}     
		\includegraphics[width=0.45\textwidth]{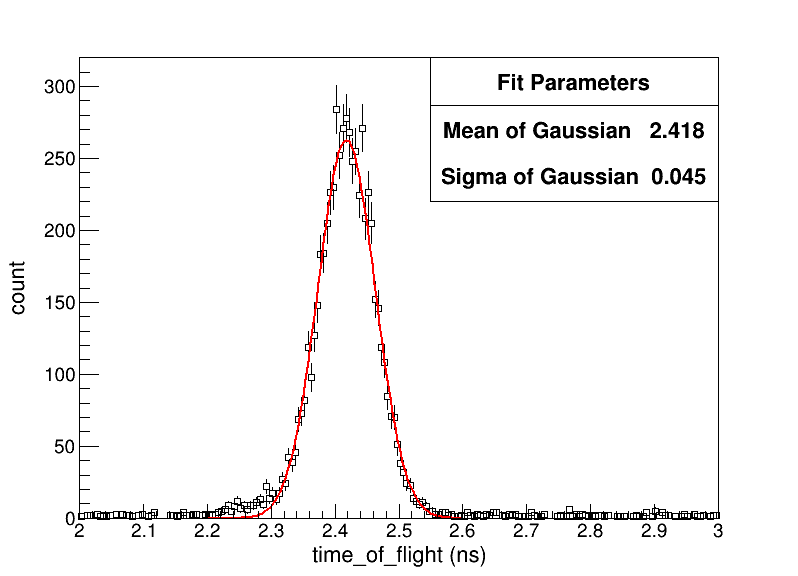}}
	\caption{Result of functional validation to the FROS. The FROS is proved to be capable of reading the fast signal of LGAD. The temporal resolution of LGAD, which is about 30 ps, can be evaluated by the $TOF_e$ distribution. Details see the text.}
	\label{fig:valid-result}
\end{figure}

Figure \ref{fig:valid-result} displays the result of the validation. 
The top plot displays a sample signal read by the FROS, with a signal width of approximately 2~ns and a rising time of around 700~ps. 
It shows that the FROS is capable of reading the fast signals of LGAD detector.
The bottom plot shows the distribution of $TOF_e$, from which the width of the $TOF_e$ distribution is evaluated using Gaussian fitting, resulting in a value of 45~ps for the standard deviation ($\sigma$). Since both LGAD detectors are identical, their temporal resolutions are expected to be the same. Therefore, the temporal resolution of the LGAD is estimated to be approximately 30~ps.

\section{Analysis and Result}
\label{sec:ana}

\subsection{Analysis to the Experiment Data}

\begin{figure}[htbp]
\centering
\includegraphics[width=.48\textwidth]{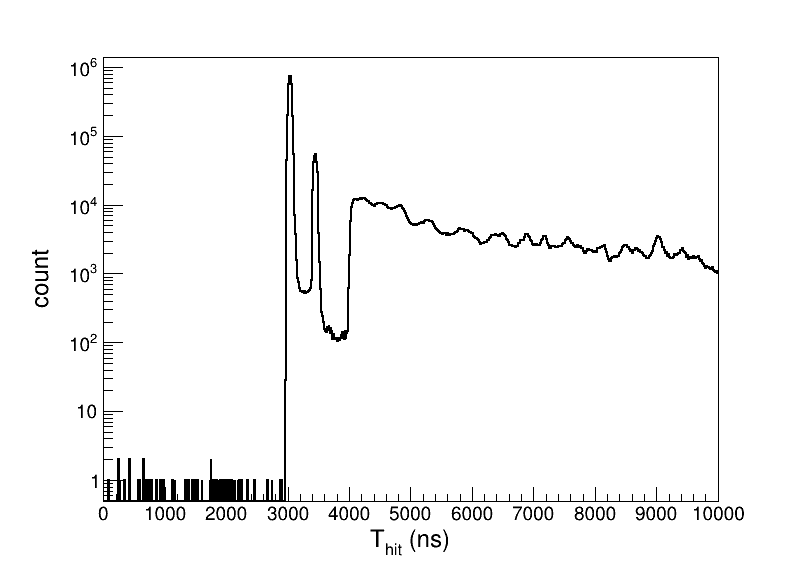}
\caption{Distribution of the arrival time of the signals referring to the $\mathrm{T_0}$ signal. The double peak in the left is the gamma flash produced in the spallation process. Two peaks are observed because the accelerator works in double bunch mode. \label{fig:gammaflash}}
\end{figure}

The main objective of this paper is to evaluate the performance of the LGAD detector when utilized as a ZDD. 
To achieve this, the cross section of the $^6Li(n,T)\alpha$ reaction is measured for neutrons with varying energies, employing the Back-n beamline at CSNS and the experimental setup described earlier.
Given the thickness of the LGAD detector, only a small fraction of the signals detected are expected to originate directly from neutrons. 
Instead, the majority of the signals are anticipated to be generated by the secondary particles produced in the $^6Li(n,T)\alpha$ reaction, namely the alpha particles and tritium.
When a secondary particle interacts with the detector, the FROS continuously records a signal, along with the time elapsed since the last $\mathrm{T_0}$ signal was received. 
The time of the signal ($\mathrm{T_1}$) is determined using a 50\% constant fraction discrimination method. 
$\mathrm{T_1}$ represents the time at which the secondary particle is detected by the LGAD, and consequently, the time at which the neutron strikes the lithium target, given their close proximity.
Figure~\ref{fig:gammaflash} illustrates a portion of the distribution of time differences between $\mathrm{T_1}$ and $\mathrm{T_0}$. 
Notably, two peaks are observed on the left side of the plot, which correspond to the gamma flash produced during the spallation process. 
These gamma particles travel faster than the neutrons and are consequently detected at the very beginning.
The presence of the gamma flash allows us to calibrate the time of flight of the neutrons, and subsequently calculate the energy of neutrons. 

By analyzing the arrival time of the gamma flash, we can determine the time at which the spallation process occurs ($t_{spa}$) using equation \ref{eq:eq1}.

\begin{equation}
    L_{exp} = c \cdot (t_{\gamma} - t_{spa})
    \label{eq:eq1}
\end{equation}

Where c denotes the light speed.
$L_{exp}$ is the distance from the center of the spallation target to the LGAD detector and has been accurately measured to be 77.0~meters. 
$t_{\gamma}$ represents the position of the first gamma flash peak. 
Then, the time of neutron flight $TOF_n$ will be calculated with eq.\ref{eq:eq2}.

\begin{equation}
    TOF_n = T_1-t_{spa}
    \label{eq:eq2}
\end{equation}

The kinetic energy of a neutron, denoted as $E_n$, can be calculated using equations \ref{eq:eq3} and \ref{eq:eq4}.

\begin{equation}
    E_n = M_n\cdot (\frac{1}{\sqrt{1-\beta ^2}} -1)
    \label{eq:eq3}
\end{equation}

\begin{equation}
    \beta = \frac{L_{exp}}{TOF_n\cdot c}
    \label{eq:eq4}
\end{equation}

\begin{figure}[h]
\centering
\includegraphics[width=.48\textwidth]{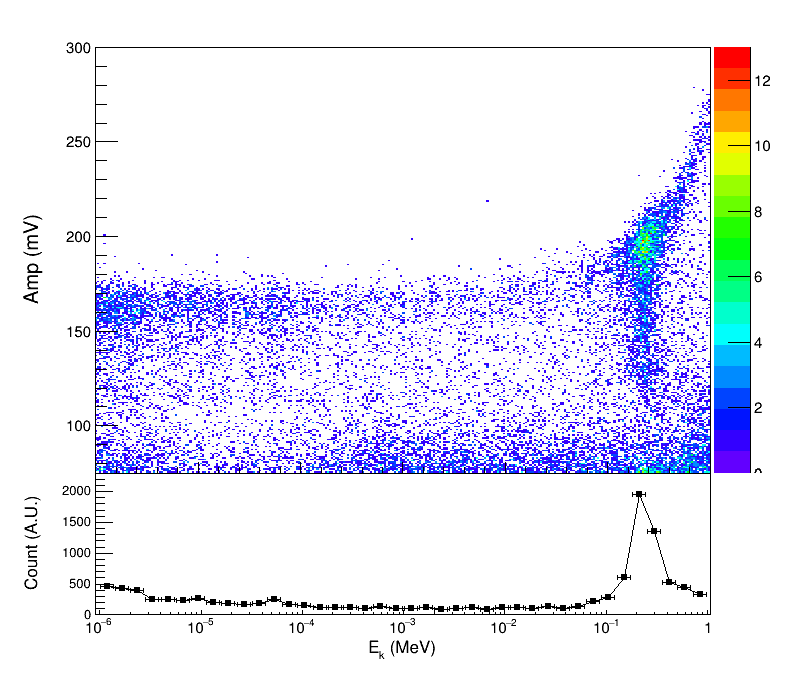}
\caption{The distribution of the reconstructed neutron kinetic energy and the amplitude of the signal. An event band is observed at the top of the plot, which corresponds to the group of T and $\alpha$. 
\label{fig:2D}} 
\end{figure}

Figure \ref{fig:2D} represents a two-dimensional distribution of reconstructed neutron kinetic energy and the amplitude of the signal generated by the secondary particles of the $^6Li(n,T)\alpha$ reaction. 
Its x-projection plot beneath displays the distribution of neutron kinetic energy, from which an obvious resonant peak can be observed. 
The amplitude of the signal corresponds to the energy deposition in the detector
An event band is observed at the top of the plot, distinct from the lower backgrounds. 
This band corresponds to the group of tritium (T) and alpha particles ($\alpha$), as the neutron and gamma background usually deposit few energy in the detector.
By selecting events within this band, the measured cross-section of the $^6Li(n,T)\alpha$ reaction can be calculated. 
This is done by correcting the neutron kinetic energy spectrum using the standard spectrum of back-scattered neutron energy, as illustrated in Figure \ref{fig:Ek}.

\begin{figure}[]
\centering
\includegraphics[width=.5\textwidth]{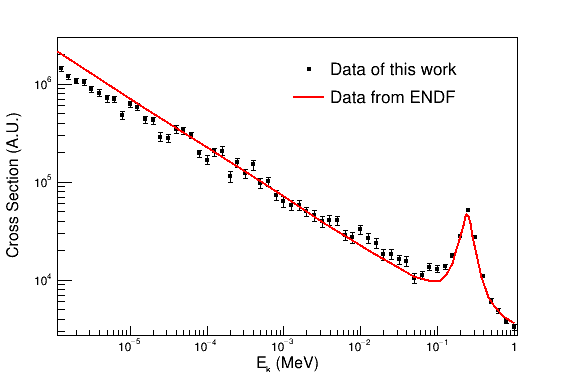}
\caption{Measured cross section of $^6Li(n,T)\alpha$ (Black dots) and data from ENDF database (Red line). The resonance peak of $^6Li$ and neutron around 200 keV can be observed on the plot. \label{fig:x-sec}}
\end{figure}

The corrected spectrum of the $^6Li(n,T)\alpha$ cross-section is depicted in Figure \ref{fig:x-sec}. 
This measured cross-section is then compared to the standard data obtained from the ENDF database. 
Notably, both data-sets exhibit agreement within the neutron energy range spanning from 100~eV to 1~MeV, although there is a disagreement for energies below 100~eV.
This discrepancy can be attributed to the fact that the experiment described in this paper was carried out simultaneously with other detector tests, which were positioned in front of our detector installation on the beamline. 
As low-energy neutrons have a higher probability of being absorbed or scattered away, fewer low-energy neutrons reached the LGAD detector and were detected than originally expected.

\subsection{Evaluation to the Anti-irradiation Performance}

The anti-irradiation performance of LGAD holds significant importance in the application of a ZDD and impacts the reliability and accuracy of data measurements in neutron-induced processes. 
Therefore, ensuring a strong anti-irradiation capability is crucial for the successful utilization of LGAD in ZDD applications. 
Reference \cite{LGAD-irr} has reported the excellent anti-irradiation performance of LGAD under the irradiated neutron flux in level of $10^{15} \mathrm{n_{eq}/cm^{2}}$ from the view of detector design.
In this paper, an additional method is proposed to evaluate this performance based on the perspective of detector application.
To evaluate the anti-irradiation performance of the LGAD, the running information of the experiment is investigated.
In total 20.64~h was the detector irradiated for on the Back-n beamline, which corresponds to a 1~MeV equivalent neutron flux of $1.10 \times 10^{10}$ passing through the detector. 
% 8.83e9 x 0.02135 / 0.01711
The anti-irradiation performance of the LGAD is evaluated based on the amplitude of charged particles ($\alpha$ or T) generated by neutrons with kinetic energy below 0.1~MeV. 
Within this energy range, the produced charged particles should exhibit nearly mono-kinetic energy characteristics.
If the LGAD demonstrates good anti-irradiation properties, the amplitude of the signal, which essentially corresponds to the kinetic energy, should follow a Gaussian distribution and remain stable throughout the experiment.

\begin{figure}[h]
    \centering
    \subfigure[Comparison to the amplitude distributions]{    
		\label{fig_s}     
	    \includegraphics[width=0.49\textwidth]{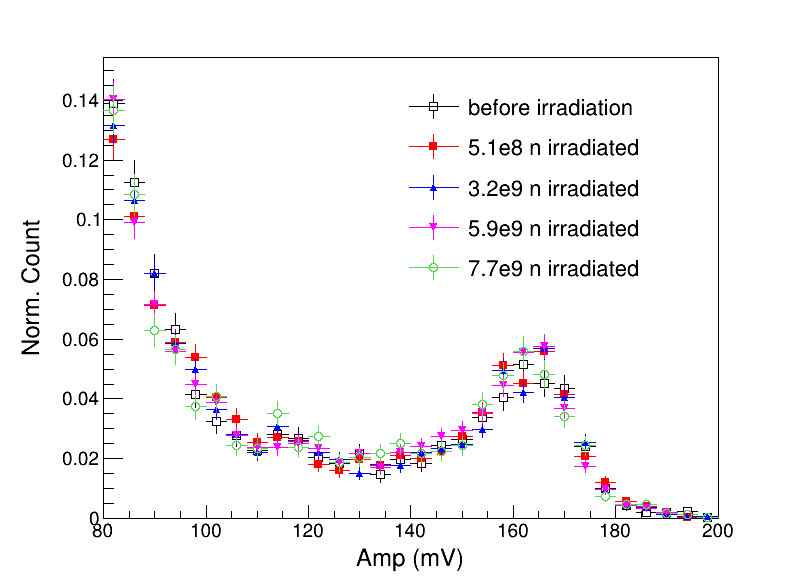}}
    \subfigure[The mean values of the distributions]{    
		\label{fig_t}     
		\includegraphics[width=0.49\textwidth]{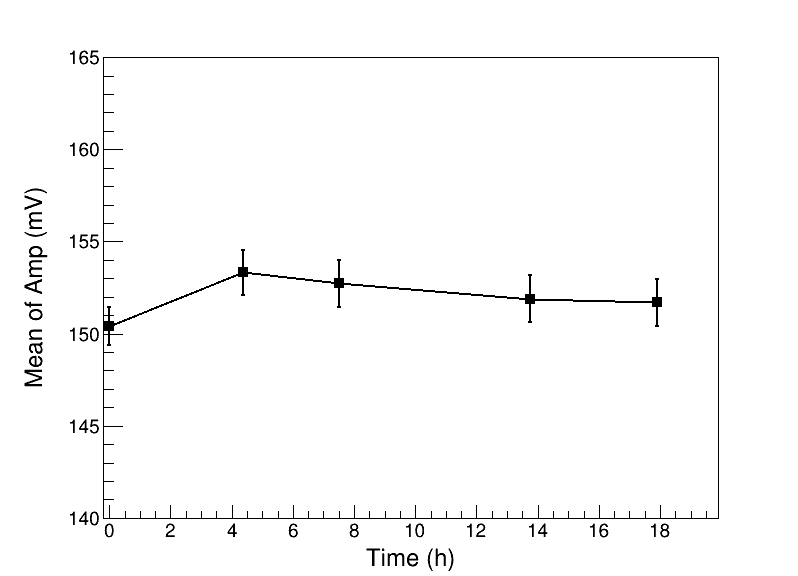}}
	\caption{Evaluation to the anti-irradiation performance of the LGAD. The peak $>$ 120~mV in (a) corresponds to the signal of $\alpha$ and T. The mean of the amplitude peak changing with the time irradiated is illustrated in (b). The LGAD exhibits good anti-irradiation performance since the means generally remain constant throughout the experiment.}
	\label{fig:irr}
\end{figure}

Figure \ref{fig:irr} (a) illustrates the amplitude distributions under different amount of neutron flux irradiation throughout the experiment.
The peaks observed when the amplitude exceeds 120 mV are mainly contributed by the signals of $\alpha$ and T, and coincide with each others.
The mean values of these distributions are presented in Figure \ref{fig:irr} (b).
It is evident that the mean values generally remain constant throughout the experiment.
This indicates that the LGAD exhibits good anti-irradiation performance, which is a significant advantage when utilizing it as a ZDD for the neutron-induced processes data measurement.

\section{Conclusion}
\label{sec:conclusion}

The LGAD detector, with its 50~\textmu m sensitive volume thickness, 30 ps temporal resolution, low cost and excellent anti-irradiation property, is expected to be an excellent candidate for the use of a ZDD in neutron-induced process data measurements. 
This paper presents an experiment that tests the performance of the LGAD when utilized as a ZDD. 
The detector is placed directly on the neutron beamline to measure the cross-section of $^6Li(n,T)\alpha$.
Comparing the results obtained in this work to the ENDF standard database, a relatively consistent outcome is observed, particularly in the resonant peak around 0.1~MeV. 
This finding suggests that the LGAD is a viable option for the application of a zero degree detector in neutron-induced process data measurements.
This will contribute to the improvement of the design of nuclear data measurement experiments, especially for that to be conducted on a short distance-of-flight neutron beamline considering its excellent temporal resolution.

%\section*{Acknowledgment}

%\section*{References}

%\def\refname{\vadjust{\vspace*{-1em}}} %Please don't do this in a real paper.

\end{document}